\documentclass[journal,twoside,web]{ieeecolor2}

\usepackage{generic}
\usepackage{cite}
\usepackage{amsfonts}
\usepackage{amsmath}
\usepackage{amssymb}
\usepackage{bm}
\usepackage{mathtools}
\usepackage{times}
\usepackage{bbm}
\usepackage{algorithm}
\usepackage{algorithmic}
\usepackage[dvipsnames]{xcolor}
\usepackage{graphicx}
\usepackage{textcomp}
\usepackage{orcidlink}
\usepackage{silence}
\usepackage{subcaption}
\usepackage{booktabs}
\usepackage{hyperref}
\hypersetup{
colorlinks=true,
linkcolor=blue,
filecolor=blue,      
urlcolor=blue,
citecolor=blue
}
\usepackage{tcolorbox}

\definecolor{lightgray}{gray}{0.9}

\newcommand{\fref}[1] {Fig.~\ref{#1}}

\newcommand{\cref}[1] {Chapter \ref{#1}}
\newcommand{\sref}[1] {Sec.~\ref{#1}}

\newcommand{\tabref}[1] {Table~\ref{#1}}
\newcommand{\eref}[1] {(\ref{#1})}

\newcommand{\xmath}[1] {\ensuremath{#1}\xspace}

\newcommand{\blmath}[1] {\xmath{\bm{#1}}}

\newcommand{\st}{\quad \text{s.t.} \quad}

\let\originalPi=\Pi 
\renewcommand{\Pi}{\blmath{\originalPi}}

\def\BibTeX{{\rm B\kern-.05em{\sc i\kern-.025em b}\kern-.08em
    T\kern-.1667em\lower.7ex\hbox{E}\kern-.125emX}}
\begin{document}
\title{Behavior Score Prediction in Resting-State Functional MRI by Deep State Space Modeling}
\author{Javier Salazar Cavazos\orcidlink{0009-0009-1218-9836}, Graduate Student Member, IEEE, \\ Maximillian Egan, Krisanne Litinas, Benjamin Hampstead\orcidlink{0000-0003-2717-6375}, and Scott Peltier\orcidlink{0000-0002-6690-9757} 
\thanks{This work was supported by the U.S. National Institutes of Health (NIH) under Grant R21 AG082204, and by the National Institute on Aging (NIA) via Michigan
Alzheimer’s Disease Research Center under grants P30 AG072931 and R35 AG072262. }
\thanks{Javier Salazar Cavazos is with 
the Electrical and Computer Engineering (ECE) Department, University of Michigan, Ann Arbor, MI 
48109 USA (e-mail: javiersc@umich.edu); Maximillian Egan, Krisanne Litinas, and Scott Peltier are with the Functional MRI Laboratory, Departments of Radiology and Biomedical Engineering (BME), University of Michigan, Ann Arbor, MI 48109 USA (e-mail: mkegan@umich.edu; klitinas@umich.edu; spelt@umich.edu); Benjamin Hampstead is with 
the Michigan Alzheimer Disease Research Center (MADRC) and Research Program on Cognition and Neuromodulation Based Interventions (RPCNBI), Departments of Psychiatry and Neurology, University of Michigan, Ann Arbor, MI 48109 USA (e-mail: bhampste@umich.edu).}}

\maketitle

\begin{abstract}
Early clinical assessment of Alzheimer’s disease relies on behavior scores that measure a subject's language, memory, and cognitive skills. On the medical imaging side, functional magnetic resonance imaging has provided invaluable insights into the neural pathways underlying Alzheimer's disease. While prior studies have used resting-state functional MRI by extracting functional connectivity matrices, these approaches neglect the temporal dynamics inherent in functional data. In this work, we present a deep state space modeling framework that directly leverages blood-oxygenation-level-dependent time series to learn a sparse set of brain regions for predicting behavioral scores. Our model extracts temporal features that encapsulate nuanced patterns of intrinsic brain activity, thereby enhancing predictive performance compared to traditional connectivity methods. We identify specific brain regions most predictive of cognitive impairment by analyzing data from the Michigan Alzheimer's Disease Research Center, providing new insights into the neural substrates of early Alzheimer's pathology. These findings have important implications for the possible development of risk monitoring and intervention strategies in Alzheimer’s disease.
\end{abstract}


\section{Introduction}
\label{sec:intro}

\begin{figure*}
    \centering
    \includegraphics[width=0.98\textwidth]{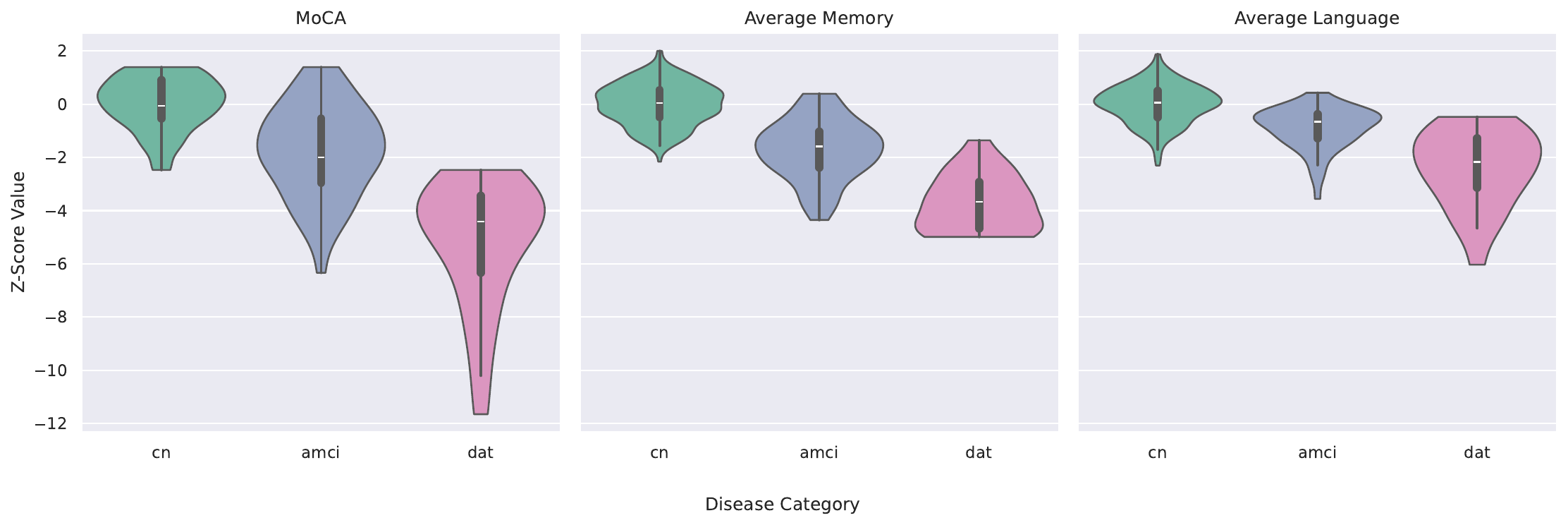}
    \caption{Violin plots illustrating the distribution of behavioral scores in z-score space across different disease categories.}
    \label{fig:moca_distribution}
\end{figure*}

\subsection{Background}
Functional magnetic resonance imaging (fMRI), which consists of a series of volumetric brain MRI scans, has been explored in various contexts to provide insights into brain function \cite{fmri_apps}. The signal acquired, namely blood-oxygen-level-dependent (BOLD) data, is used as a proxy for neural activity that impacts hemodynamics at the voxel level. Resting-state fMRI (rs-fMRI) is a neuroimaging technique that measures spontaneous brain activity by detecting regional differences in cerebral blood flow while a person is not performing any explicit task under the scanner. It is important because it reveals patterns of functional connectivity between brain regions, providing insights into brain organization and various neurological or psychiatric conditions \cite{fmri}. These patterns indicate correlated brain activations that can reflect physiological changes in the brain \cite{rsfmri1, rsfmri2}. Recent work has focused on identifying neurological diseases using machine learning models based on functional connectivity \cite{disease1, disease2}, a technique that measures correlations in BOLD activity among brain regions. Further, various studies have shown that the default mode network (DMN), a collection of brain regions active when a person is not performing a specific task in rs-fMRI, constitutes a baseline state that involves self-referential processing and internally directed cognition \cite{dmn1, dmn2}. In this work, we explore whether the default mode network plays a significant role in cognitive impairment by predicting behavior scores. 

Alzheimer's disease (AD) is a progressive neurodegenerative disease that often leads to dementia, characterized by loss of cognitive and memory function. Healthy patients are known as cognitively normal (CN), and clinicians have classified Alzheimer’s disease into three distinct stages: the preclinical stage, an intermediate phase known as mild cognitive impairment (MCI), and, in the later stages, dementia of the Alzheimer's type (DAT) \cite{ad_stages}. Preclinical AD affects the brain years before any diagnosis is made, and so there is a need to study brain changes in the early stages to aid detection and treatment methods. Unfortunately, there are often confounders or mixed pathologies, e.g., depression or Parkinson's disease \cite{confounders}, that complicate the diagnosis process. This places high importance on adaptable biomarkers that can differentiate overlapping pathologies such as fMRI, positron emission tomography (PET), and others. The first and most viable diagnostic marker, meaning the cheapest and easiest to implement, is one that does not require any imaging, such as a written cognitive exam that tests for things such as memory, language, and general cognition. The primary cognitive measure, also called a behavior score, analyzed in this work is the Montreal Cognitive Assessment (MoCA), which is designed for screening MCI subjects, and has been shown to be superior to other exams such as the Mini-Mental State Examination (MMSE) \cite{moca}. The MoCA exam includes tasks in several areas, such as short-term memory recall, delayed recall, visual-spatial skills, executive function, sustained attention, language, and time-place orientation. MoCA can discriminate between MCI and CN subjects with an area under the curve (AUC) of 0.86, indicating the practicality of this metric for early diagnosis \cite{moca}. Additionally, in this work, we also compute average memory and language metrics for the prediction task. Incorporating these average subcategories, in addition to the MoCA metric, provides finer, detailed information on the subject's cognitive performance. 

\subsection{Related Works}

The intersection between behavior score prediction and imaging modalities is not entirely unprecedented. For example, the brain-behavior relationship has been studied in PET imaging using standardized uptake value ratio (SUVR) features \cite{mocapet}. More closely to this work, the association with MoCA and depression subjects has been studied in an rs-fMRI context \cite{mocadepression} by using the functional connectivity predictive modeling (CPM) method \cite{cpm}. In terms of behavior score prediction and AD subjects, Ref. \cite{mocarsfmri} predicts the ADAS-Cog \cite{adascog} metric using CPM on subjects spanning the AD spectrum. Further, Ref. \cite[Chapter 2]{michelle_thesis} does something similar with functional connectivity data using the CPM method in a cohort of purely MCI subjects. The Pearson correlation coefficient (R) for MoCA prediction is 0.07 in Ref. \cite[Chapter 2]{michelle_thesis} and 0.15 in Ref. \cite{mocadepression}, indicating a fairly weak correlation for MoCA prediction in rs-fMRI. It is worth noting that this is not a fair comparison, as these works use different rs-fMRI datasets and methodologies. However, these works consider only functional connectivity data, which collapses the time dimension and examines only brain region interactions. To the best of the authors' knowledge, there are no works that tackle behavior score prediction on the BOLD timeseries data itself, which may provide a richer context for finding more optimal predictions. Furthermore, these works do not investigate cognitive subcategories such as memory and language in this prediction problem, only broad metrics such as MoCA.

\subsection{Contributions \& Motivation}
In recent brain stimulation research, Ref. \cite{brainstim} used high-definition transcranial direct current stimulation (HD-tDCS) to target the left ventrolateral prefrontal cortex exclusively in individuals with mild cognitive impairment (MCI). This intervention was administered in a double-blind protocol alongside either mnemonic strategy training or an autobiographical recall control condition over five consecutive daily sessions. The findings demonstrate pronounced neurophysiological effects during associative memory encoding in the experimental group relative to controls. These results underscore the importance of delineating specific brain regions implicated in cognitive impairment, thereby informing the potential of HD-tDCS applications in AD that extend beyond single-region stimulation paradigms. Given the recent rise of ultrasound brain stimulation \cite{ultrasound_stimulation}, that can target multiple regions beyond the brain surface, future studies can explore these techniques to improve patients' cognitive state in the AD spectrum.

In this work, we systematically compare the rs-fMRI modality for predicting behavior scores in subjects spanning the AD spectrum using MoCA and subcategory metrics. This is done by exploring both functional connectivity and multivariate timeseries methods for this task, developing a data-driven deep learning method based on state space models that outperforms the other approaches, and drawing biological insights into the brain-behavior relationship.

\begin{table*}
\centering
\resizebox{0.98\textwidth}{!}{%
\begin{tabular}{cccccccc}
\hline
Category & Total Subjects & Age & Education & Race (W/B/A) & Sex (M/F) & MoCA \\ \hline
CN   & 202 & 71.6 $\pm$ 6.1 & 16.9 $\pm$ 2.1  & 168/31/3 & 131/71 & 27.1 $\pm$ 2.1  \\ 
aMCI & 53 & 73.3 $\pm$ 6.8 & 16.1 $\pm$ 2.2 & 44/8/1   & 16/37   & 23.4 $\pm$ 3.5  \\ 
DAT  & 26 & 71.9 $\pm$ 6.2 & 15.9 $\pm$ 2.6 & 25/0/1   & 12/14  & 16.6 $\pm$ 5.0 \\ 
\end{tabular}%
}
\caption{Distribution of subjects in the MADRC dataset, stratified by disease category.}
\label{tab:madc}
\end{table*}

\section{Data Acquisition \& Processing}
\label{sec:data}

\subsection{Dataset}
The Michigan Alzheimer's Disease Research Center (MADRC), in collaboration with the University of Michigan, has collected functional and structural 3T MRI single-session scans, along with behavioral score metrics, for use in this project as part of the National Institute on Aging (NIA) grants P30 AG072931 and R35 AG072262. All participants provided written informed consent, and study activities were approved by the respective parties. Particularly, this study has been approved by the University of Michigan Human Research Activation Committee on February 13, 2026, with Institutional Review Board (IRB) submission number HUM00000382. Resting-state fMRI scans were completed with eyes open with a fixation cross. The 3T rs-fMRI data is acquired with a multi-band (MB) echo-planar imaging (MB-EPI) pulse sequence with MB factor of 6 leading to a 30ms echo time (TE), 0.8s repetition time (TR), a flip angle of $52^{\circ}$, a 2.4mm spatial resolution, 60 total slices, and 570 total time samples. 

The MADRC subject population consists of 281 subjects across all disease category labels. Generally, a subject can be cognitively normal (CN), have mild cognitive impairment (MCI), or dementia of the Alzheimer's type (DAT). MCI subjects that have memory-related impairment have the amnestic MCI (aMCI) label, while subjects that exhibit cognitive impairment (without memory-related impairment) have the non-amnestic MCI (naMCI) label. In the medical literature, this impairment grouping is common since aMCI subjects are more likely to develop into DAT subjects in the future. All MADRC MCI subjects used in this study are amnestic. For additional details about the subject population, see \tabref{tab:madc} for demographic characteristics. Table acronyms are listed as follows: CN = cognitively normal, aMCI = amnestic mild cognitive impairment, DAT = Dementia of the Alzheimer's Type, W = White/Caucasian, B = Black/African American, A = Asian, M = Male, and F = Female. The behavior score distribution among subjects is illustrated in \fref{fig:moca_distribution} by disease category and behavior score subcategories.

\subsection{Behavior Score Metrics}

As stated earlier, the MoCA score is the main cognitive metric used in this work. Additionally, memory and language submetrics are computed to provide a deeper understanding of cognitive impairment. In greater detail, a composite memory metric was calculated, averaging the z-scores of the UDS Benson complex figure recall \cite{uds}, CRAFT story delayed recall \cite{craft}, and CRAFT delayed verbal \cite{craft} behavioral tasks. A composite language metric was calculated, averaging the z-scores of total numbers of named animals \cite{uds}, total number of named vegetables \cite{uds}, and the Multilingual Naming Test (MINT) exam \cite{mint}. All metrics are z-score normalized using healthy population data.

\subsection{Preprocessing}
\label{sec:preprocess}
Firstly, the raw data is processed by converting it to the standard BIDS format \cite{bids} for use in downstream software packages. Secondly, the application fmriprep \cite{fmriprep} handles most standard tasks such as N4 bias field correction, skull stripping, normalization to the MNI152 linear space provided by the TemplateFlow archive \cite{templateflow}, head-motion estimation, susceptibility distortion correction using fieldmaps, and z-score normalization of each voxel's timeseries data. 


Lastly, the BOLD data is parcellated into 264 regions of interest (ROIs) using the Power atlas \cite{power_atlas}, with 8 additional ROIs we manually included ourselves to augment the atlas, since some areas of the amygdala and hippocampus may not be sufficiently covered, giving a total of 272 ROIs used in this project. The MNI coordinates and labels for these additional ROIs are included in \tabref{tab:roi}. Please see \href{https://github.com/javiersc1/NeuroMamba}{github.com/javiersc1/NeuroMamba} for additional information regarding our specific preprocessing pipeline.

\begin{table}
\centering
\resizebox{0.66\columnwidth}{!}{%
\begin{tabular}{ccc}
\hline
ROI & MNI (x, y, z) & Label   \\ \hline
265          & (-24, -4, -20)         & Left Amygdala     \\ 
266          & (24, -2, -20)          & Right Amygdala    \\ 
267          & (28, -12, -20)         & Right Hippocampus \\ 
268          & (30, -24, -12)         & Right Hippocampus \\ 
269          & (30, -39, -3)          & Right Hippocampus \\ 
270          & (-29, -12, -22)        & Left Hippocampus  \\ 
271          & (-30, -24, -12)        & Left Hippocampus  \\ 
272          & (-29, -38, -4)         & Left Hippocampus  \\ 
\end{tabular}%
}
\caption{MNI coordinates and anatomical labels for additional regions of interest (ROIs) incorporated alongside the 264 ROIs defined in the Power atlas.}
\label{tab:roi}
\end{table}
\section{Formulation \& Related Methods}

\subsection{Problem Formulation}
For each subject $i$, a 3D volume is acquired over time to get BOLD data $V^{i} \in \mathbb{R}^{T \times S_X \times S_Y \times S_Z }$ where $T$ represents time and $(S_X,S_Y,S_Z)$ are spatial dimensions. After the preprocessing outlined in \sref{sec:data}, the timeseries data becomes $X^i \in \mathbb{R}^{T \times B} $ where $B=272$ refers to the brain region ROIs and $T=540$ represents time samples. This timeseries data $X^i$ will be the main focus of this work. 

The goal then is to find some function $g(\cdot)$ that maps the given input, either $X^i$ or some derivative of it, to the output $s^i \in \mathbb{R}^3$ where $s^i$ represents the MoCA, memory, and language metrics for the $i$th subject. All subjects' scores are collected into a single matrix $S = [\hspace{1mm} s^1, \ldots, s^N \hspace{1mm}]$ where $N$ denotes the number of samples. In this section, different methods are introduced that extract features from $X^i$ to be used in a ridge regression model setting as discussed in \sref{ridge_regression}.

\subsection{Kernel Ridge Regression (KRR)}
\label{ridge_regression}
In standard ridge regression \cite{ridge_regression}, the objective is to minimize the loss function with a regularization term to prevent overfitting by solving
\begin{equation}
    \min_{W,B} \underbrace{\frac{1}{2} \| S - FW -B \|_{\text{F}}^2}_{\hidewidth \text{data fidelity} \hidewidth} + \lambda \underbrace{\| W \|_\text{F}^2}_{\hidewidth \text{prevent large values} \hidewidth}
    \label{eq:rr}
\end{equation}
where $\|\cdot\|_F$ is the Frobenius norm, $W$ are the regression weights, $B$ is the bias term, and $\lambda > 0$ is the regularization parameter. The input variable $F$ consists of some features of the timeseries data $\mathcal{X} = \{\hspace{1mm }X^1, \ldots, X^N \hspace{1mm} \}$. In \eref{eq:rr}, $g(F) = F W+B$ is the linear model assumption for $F$. To extend this approach, kernel ridge regression (KRR) \cite{kernel_ridge} replaces the inner product of feature vectors with a kernel function $k(\cdot,\cdot)$, implicitly mapping inputs into a higher dimensional space defined by the kernel. The kernel matrix $K \in \mathbb{R}^{N \times N}$ is computed such that $K_{i,j} = k(F_{i,:}, F_{j,:})$. The KRR model optimizes the following objective
\begin{tcolorbox}
\centering
\textbf{Kernel Ridge Regression (KRR)}
\begin{equation}
    \min_{W,B} \frac{1}{2} \| S - KW-B \|_\text{F}^2 + \lambda \| K W \|_{\text{F}}^2.
    \label{eq:krr}
\end{equation}
\end{tcolorbox}

In our experiments, we use the radial basis function (RBF) kernel, also called the Gaussian kernel, defined as
\begin{equation}
    k(F_{i,:}, F_{j,:}) = \exp(-\gamma \| F_{i,:} - F_{j,:} \|_2^2)
\end{equation}
where $\gamma>0$ is the kernel hyperparameter that controls the width or spread of the kernel. A larger $\gamma$ means the kernel declines rapidly with distance focusing more on close neighbors while a smaller $\gamma$ makes the kernel smoother and more global. The RBF kernel is the most wildly used kernel function in machine learning because of its versatility and strong performance on a wide variety of tasks \cite{kernels}. By using it, the model can capture nonlinear relationships between input features $F$ and score metrics $S$, thereby improving predictive accuracy. From \eref{eq:krr}, it should be clear that different weights and bias terms are computed for each behavior score metric given the same input feature matrix $F$. 

\subsection{Connectivity-Based Methods}
\label{connectivity}
\subsubsection{Functional Connectivity (FC)}
One can generate functional connectivity data by forming the Pearson product-moment correlation coefficient matrix using $X^i$. Let $C^i \in \mathbb{R}^{B \times B}$ represent the functional connectivity matrix. Mathematically, this is computed by 
\begin{tcolorbox}
\centering
\textbf{Functional Connectivity (FC)}
\begin{equation}
    \forall j,k \in \mathcal{B}, \quad C^i_{j,k} = \frac{\text{cov}(X^i_{:,j},X^i_{:,k} )}{\sigma({X^i_{:,j}}) \hspace{1mm} \sigma(X^i_{:,k})}
    \label{eq:fcm}
\end{equation}
\end{tcolorbox}
\noindent where $\mathcal{B}$ is the set of brain regions, $\text{cov}(\cdot)$ is covariance, and $\sigma(\cdot)$ is standard deviation. Since the functional connectivity matrix is symmetric, the upper half is extracted by the linear operator $\mathcal{A}$ such that $\tilde{C}^i = \mathcal{A}(C^i) \in \mathbb{R}^{\tilde{B}}$ is a vector whose elements are the brain region correlations and where $\tilde{B} = B(B-1)/2$. Let $F = [ \tilde{C}^1, \ldots, \tilde{C}^N ]$ be the matrix used in ridge regression by solving \eref{eq:krr}.

\subsubsection{Connectome Predictive Modeling (CPM)}
Connectome predictive modeling (CPM) \cite{cpm} is an approach to relate functional network strength to specific metrics. CPM has been applied in various studies of connectivity such as predicting attention performance \cite{cpm1}, individual identification \cite{cpm2}, and fluid intelligence \cite{cpm3}. The method works by correlating connectivity data and behavioral measures. The resulting correlations are thresholded at a given level to determine significance. Next, the connectivity values at significant edges are summed separately for positive and negative correlations. This generates a ``brain score'' that is used in a linear regression setup.

Let $\tilde{C}^i = \mathcal{A}(C^i)$ represent the vectorized functional connectivity for the $i$th subject by applying \eref{eq:fcm}. Let $\mathcal{C}_j$ represent a vector that consists of the $j$th edge values across the population. Let $S_k$ represent a vector of the $k$th behavioral measure across the population. Then, the CPM method works by first generating a mask vector $M^k \in \mathbb{R}^{\tilde{B}}$ that selects the most significantly correlated edges, and summing the edges values for each subject. Mathematically, this can be expressed as
\begin{tcolorbox}
\centering
\textbf{Connectome Predictive Modeling (CPM)}
\begin{align}
    M^k_{j} &= \mathcal{T}^{+/-}( \text{corr}(\mathcal{C}_j, S_{k}) ), \hspace{1mm} \forall j \in \{1, \ldots, \tilde{B}\} \nonumber \\
    f^i_k &= \langle \tilde{C}^i \odot M^k, \mathbbm{1}_{\tilde{B}} \rangle
\end{align}
\end{tcolorbox}
\noindent where $\text{corr}(\cdot,\cdot)$ is the correlation function, $\mathcal{T}^{+/-}(\cdot)$ is the operator that constructs a binary mask that comes from the most significantly correlated positive or negative edges, $\odot$ is element-wise multiplication, $\mathbbm{1}_{\tilde{B}}$ is the ones vector of size $\tilde{B}$, and $f^i_k$ is the brain score feature extracted for the $i$th subject associated with the $k$th behavior score. The matrix $F = [\hspace{1mm} f^1, \ldots, f^N \hspace{1mm}]$ is the input feature matrix used in ridge regression by solving \eref{eq:krr}.

\subsection{Model-Based Timeseries Methods}
\label{timeseries}
\subsubsection{Individual Independent Component Analysis (I-ICA)}
Independent component analysis (ICA) \cite{ica} has become a fundamental tool in fMRI research for decomposing BOLD timeseries data into spatially independent components representing distinct functional networks \cite{ica1}. ICA enables the identification of both task-based and resting-state networks without prior knowledge of temporal profiles, making it particularly valuable for exploratory analyses. To give a few examples, some common applications are the isolation of resting-state networks such as the default mode network \cite{ica1}, characterization of brain connectivity changes in neurological and psychiatric conditions \cite{ica2}, and the separation of physiological noise or motion artifacts from true neural signals \cite{ica3}. In our experiments, we use ICA to extract features from the multivariate timeseries data $X^i$.

ICA, similar to principal component analysis (PCA), is a method where one finds a low-dimensional feature space that consists of independent sources/components and mixing coefficients that maximizes the mutual information between the feature space and ambient space. The model assumes that $X^i=M^iE^i$ where $M^i$ are the mixing coefficients and $E^i$ are the independent sources/components. The goal is to find the inverse mapping $W^i=(M^i)^{-1}$ so given the original data $X^i$, one finds the components by unmixing, i.e., $E^i=W^iX^i$. Mathematically, given timeseries data $X^i \in \mathbb{R}^{T \times B}$ that has zero mean and unit covariance, ICA is associated with the following objective
\begin{tcolorbox}
\centering
\textbf{Individual Independent Component Analysis (I-ICA)}
\begin{equation}
    \min_{W^i} \underbrace{\frac{1}{2} \| a(W^i X^i) \|_F^2}_{\hidewidth \text{sparsity regularizer} \hidewidth} \st \underbrace{W^i (W^i)'=I}_{\hidewidth \text{orthonormal sources} \hidewidth}
    \label{eq:iica}
\end{equation}
\end{tcolorbox}
\noindent where $W^i$ contains the inverse mixing coefficients, $W^i X^i$ are the components, $a$ is a nonlinear convex function such as $\log(\cosh(\cdot))$, and $W^i (W^i)'=I$ is the independent constraint. This is done for each subject independently to extract the components, and we denote this version as individual ICA (I-ICA). In this work, the components are features used in ridge regression by solving \eref{eq:krr} on the input matrix $F = [\hspace{1mm} \hat{W}^1 X^1, \ldots, \hat{W}^N X^N \hspace{1mm} ]$ where $\hat{W}^i$ is the solution to solving \eref{eq:iica} for each $i$th subject. We denote this version as Individual ICA (I-ICA).

\subsubsection{Group Independent Component Analysis (G-ICA)}
In classical ICA for fMRI, labeled here as I-ICA, each subject has components extracted independently from each other which can result in inconsistencies among multiple subjects. The classical ICA method does not generalize to draw inferences about groups of subjects since different subjects will have different time courses so it is not immediately clear how to extend the method for group data \cite{group_ica}. Thus, modifications are required to make the method more meaningful in a group context; many approaches have been developed for this in the fMRI context \cite{group_ica}. The most wildly used group ICA method is a temporal concatenation approach known as GIFT/MELODIC \cite{gift} where a matrix $X^G \in \mathbb{R}^{ TN \times B}$ is formed that consists of temporally stacked data for all subjects. This ensures common sources are found across the population. Let $\mathcal{T}_i$ represent the set of all time samples associated with subject $i$ and $\hat{W}_G$ is the result of solving \eref{eq:iica} on $X^G$. Each subject contains a submatrix $\hat{W}^i_{G} = \hat{W}_G[\mathcal{T}_i , :]$ in $\hat{W}_G$ that contains the unmixing coefficients. Mathematically, this is expressed as doing the following
\begin{tcolorbox}
\centering
\textbf{Group ICA (G-ICA)}
\begin{align}
    &\text{1. Let } X^G =  
\begin{bmatrix}
\mathbin{\text{---}} X^1_{:,1} \mathbin{\text{---}} & \ldots & \mathbin{\text{---}} X^N_{:,1} \mathbin{\text{---}} \\
\vdots & \ddots & \vdots\\
\mathbin{\text{---}} X^1_{:,B} \mathbin{\text{---}} & \ldots & \mathbin{\text{---}} X^N_{:,B} \mathbin{\text{---}}
\end{bmatrix} \nonumber \\
&\text{2. Apply ICA \eref{eq:iica} on } X^G \nonumber \\
&\text{3. Extract } \hat{W}^i_{G} = \hat{W}_G[\mathcal{T}_i , :], \hspace{1mm} \forall i \in \{1, \ldots, N\}
    \label{eq:gica}
\end{align}
\end{tcolorbox}
\noindent Similarly as before, the components $\hat{W}^i_{G} X^i$ are extracted to form $F = [\hspace{1mm} \hat{W}^1_{G} X^1, \ldots, \hat{W}^N_{G} X^N \hspace{1mm}]$ used in ridge regression by solving \eref{eq:krr} on $F$. We denote this version as Group ICA (G-ICA).

\subsubsection{Amplitude of Low Frequency Fluctuations (ALFF)}
The amplitude of low frequency fluctuations (ALFF) \cite{alff} is a metric derived from BOLD timeseries data that extracts Fourier coefficients. ALFF quantifies the intensity of activity by measuring the power spectral density of low frequency fluctuations for each brain region timeseries signal. Higher ALFF values indicate greater regional neural activity, providing insights into baseline brain function and potential alterations associated with neuropsychiatric conditions. To give a few examples, this method was used to predict frequency of migraines in individuals \cite{alff_migrane}, and identify brain regions relevant in schizophrenia subjects \cite{alff_schizo}. Mathematically, for each subject $i$, this method computes
\begin{tcolorbox}
\centering
\textbf{Amplitude of Low Frequency Fluctuations (ALFF)}
\begin{equation}
    a^i = \left[ \mathcal{M}( |\mathcal{Z}_1(\mathcal{F}(X^i_{b,:}))|^{1/2})  \text{ for } b \in \{1, \ldots, B\} \right]
    \label{eq:alff}
\end{equation}
\end{tcolorbox}
\noindent where $\mathcal{M}(\cdot)$ computes the mean, $\mathcal{Z}_1(\cdot)$ extracts the Fourier coefficients associated with frequencies 0.008-0.9Hz, $\mathcal{F}(\cdot)$ is the Fourier transform, and $|\cdot|$ computes the magnitude of the complex coefficients. The feature matrix is computed to form $F = [\hspace{1mm} a^1, \ldots, a^N \hspace{1mm}]$ and this is used as the input to ridge regression by solving \eref{eq:krr}. 


\subsection{Data-Driven Timeseries Methods}
\label{sec:related_deep}
\subsubsection{Temporal Convolutional Network (TCN)}
Temporal Convolutional Networks (TCNs) \cite{tcn} are deep models designed specifically for processing sequential data. They are based on convolutional neural networks (CNNs), which are well-known for their efficiency in processing visual data such as images. In TCNs, the convolutional layers are applied along the time dimension of the sequential data. This model is adapted for regression by changing the head to instead perform global average pooling along the time dimension and adding a linear layer to predict behavior scores from the number of channels or variates in this instance.

\subsubsection{Long Short-Term Memory (LSTM)}
Long short-term memory networks (LSTMs) \cite{lstm} are a specialized type of a recurrent neural network (RNN) designed to address the vanishing and exploding gradient problems commonly encountered in standard RNNs. By incorporating gated memory cells, LSTMs can effectively capture and utilize long-range dependencies in sequential data, making them highly successful across a variety of tasks. This is due to their use of input, output, and forget gates, which allow the model to selectively retain relevant information over extended sequences. Since their introduction, LSTMs have served as a foundational building block for many advances in sequence modeling and have inspired a range of related architectures. In our experiments, a bidirectional variant \cite{bilstm} of the LSTM model is used and denoted as ``BiLSTM''. This architecture is adapted for regression, similar to the TCN model.

\subsubsection{Patch Time Series Transformer (PatchTST)}
The Patch Time Series Transformer (PatchTST) \cite{patchtst} adapts the Transformer architecture \cite{transformer} for time series tasks by introducing a patch-based input representation. Unlike traditional approaches that operate on individual time steps, PatchTST partitions the input sequence into overlapping patches, enabling the model to capture both local and global patterns efficiently. This design leverages self-attention mechanisms to model complex temporal dependencies across multiple variables. PatchTST has achieved excellent performance on multivariate timeseries forecasting tasks \cite{patchtst}, demonstrating superior performance compared to classical Transformer models. This architecture is adapted for regression, similar to the TCN model.
\section{Proposed Method}
\label{sec:method}

\begin{figure*}[!ht]
    \centering
    \includegraphics[width=\textwidth]{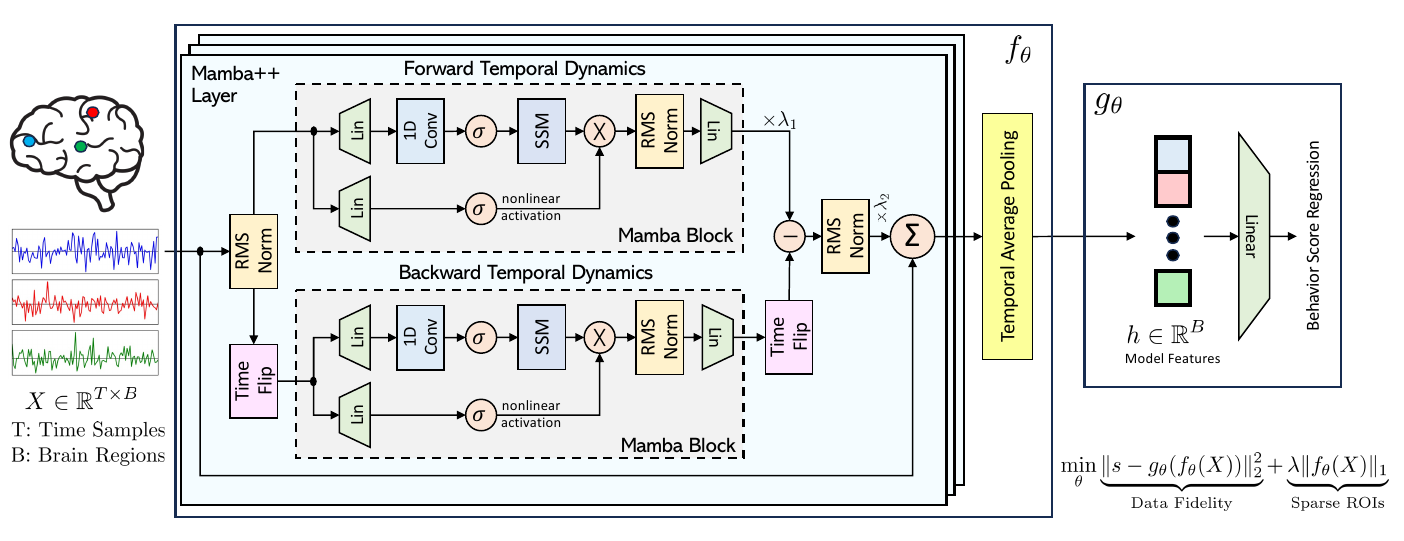}
    \caption{Overview of the proposed NeuroMamba architecture for behavior score prediction using deep state space modeling. The Mamba++ layer extracts temporal features from each brain region relevant to prediction. Temporal averaging is subsequently applied to derive a single scalar summary statistic per region, which is then processed by a linear head.}
    \label{fig:mambamodel}
\end{figure*}

\subsection{Deep State Space Models (SSMs)}
State space models (SSMs) model physical systems using state variables that track how inputs change system behavior over time and can be written in matrix form for linear, time-invariant (LTI) systems \cite{ssm}. Given an input signal $x(t) \in \mathbb{R}$, the SSM will generate an output sequence $y(t) \in \mathbb{R}$ using the state $h(t) \in \mathbb{R}^L$ by equations
\begin{align}
    h'(t) &= A h(t) + B x(t) \nonumber \\
    y(t) &= C h(t) + D x(t)
    \label{eq:ssm}
\end{align}
where $A$ is the state matrix, $B$ is the input matrix, $C$ is the output matrix, $D$ is the feed-through matrix, and $L$ is the state size. In well-known physical problems, the $\{A,B,C,D\}$ matrices are determined by differential equations or by physics-based modeling of the system. In contrast, for complex systems, the dynamics are either unknown or difficult to model, so the $\{A,B,C,D\}$ matrices are learned in a data-driven way in the deep SSM context. The first popular work that studies this deep SSM idea and addresses some important challenges, such as matrix initialization and fast parallelized computation, is the structured SSM (S4) model \cite{s4}. Since acquired data is discrete with some time step $\Delta$, the recurrence equation in \eref{eq:ssm} is rewritten using zero-order hold discretization \cite{mamba} by
\begin{tcolorbox}
\centering
\textbf{S6 Block}
\begin{align}
    \bar{A} &= \exp({\Delta A}), \quad \bar{B} = (\Delta A)^{-1} (\exp({\Delta A}) - I) \cdot \Delta B, \nonumber \\
    y_k &= C h_t, \quad h_t = \bar{A} h_{t-1} + \bar{B} x_t.
    \label{eq:s4}
\end{align}
\end{tcolorbox}
\noindent Note that $D$ is dropped since it can easily be implemented as a skip connection in a neural network context. Note that both \eref{eq:ssm} and \eref{eq:s4} apply to a single sequence; to process multiple variates, multiple independent SSM instances are used in practice.

However, since the SSM modeling in \eref{eq:ssm} and \eref{eq:s4} is time-invariant, this poses challenges for certain tasks such as noun selection in language problems. Thus, the S6 layer, named after S4 combined with selective scanning, employs a selective mechanism in which $\{A,B,C\}$ are time-varying, making the SSM content-aware \cite{mamba, mamba2}. However, since the system is no longer LTI, the update rules for $\{A, B, C\}$ can no longer be easily parallelized in FFT multiplication form. Instead, the recurrence equation is used, but is still efficiently parallelized using parallel scan methods, also known as parallel prefix sum algorithms \cite{parallel_scan}. Additionally, the S6 layer remains linear, so in practice the layer is wrapped around a ``Mamba block'' formalized by
\begin{tcolorbox}
\centering
\textbf{Mamba Layer}
\begin{align}
    H &= \sigma (\text{Conv1D}(\text{Affine}(X))), \quad Z = \sigma(\text{Affine}(X)), \nonumber \\
    Y &= \text{S6}(H), \quad \hat{Y} = \text{Affine}(Y \odot Z)
\end{align}
\end{tcolorbox}
\noindent where $X$ is the input, $\hat{Y}$ is the output, $\sigma$ is SiLU activation, $\text{S6}(\cdot)$ is the operation in \eref{eq:s4}, and $\odot$ is element-wise multiplication with the gating mechanism $Z$. This time-varying S6 layer, wrapped around nonlinear components, enables a non-LTI SSM formulation that better captures the nuances of complex systems that do not satisfy linearity or time-invariance. However, this Mamba model was primarily designed for language tasks and therefore does not leverage domain-specific knowledge, such as multivariate timeseries data. In \sref{sec:neuromamba}, modifications to the Mamba model are proposed, such as bidirectionality and differential design, to aid with modeling temporal dynamics in the timeseries data.

\subsection{NeuroMamba}
\label{sec:neuromamba}
\subsubsection{Bidirectionality}
To enhance Mamba's ability to preserve historical information over a longer time range, we generalize by introducing two Mamba blocks: one to capture forward temporal dynamics and another to capture backward temporal dynamics. This addresses the limitation of SSMs, which forget past information at longer horizons \cite{hippo}, by focusing on both recent and past information. We note that this kind of idea is not novel, as similar ideas have already been proposed for long short-term memory networks (LSTMs) and others \cite{bilstm}. The Mamba model does not have this functionality because it was designed for language tasks where the goal is to auto-regressively predict the next token in a left-to-right fashion. However, in this fMRI application, one is more interested in learning from temporal dynamics to predict behavior scores, and because this is not a model for real-time use, bidirectionality enables a more predictive model in this offline context.

\subsubsection{Differential design}
Sequence models such as Transformers and RNNs tend to over-allocate attention to irrelevant context, especially for multi-needle retrieval language tasks where the goal is to answer questions embedded in a pile of documents. Recent works have focused on differential attention \cite{dif_transformer} to calculate the difference between two attention mechanisms, and this was found to cancel noisy intermediate representations, similar to differential amplifiers in electrical engineering. This led to advantages in key areas such as information retrieval, hallucination mitigation, in-context learning, and the reduction of activation outliers \cite{dif_transformer}. This idea has been recently extended to RNNs and SSMs \cite{dif_mamba}. We incorporate this idea because of similarities to our problem, e.g., finding a few variates embedded within the entire list that are most impactful for prediction, similar to multi-needle retrieval. Because this differential design never performed worse than the baseline model \cite{dif_transformer}, we integrate bidirectionality to instead compute the learnable scaled difference rather than simply adding the two Mamba blocks. 

\subsubsection{Small Batch Regularization (SBR)}
Neural networks with strong performance are almost always over-parameterized, which is why they are called ``deep''. Despite explicit regularization to avoid overfitting, some studies have found that the implicit regularization of small-batch training tends to converge to minima with better generalization performance, regardless of whether explicit regularization is used \cite{smallbatch1, smallbatch2, smallbatch3}. Specifically, with limited training samples, the regularization effect remains strong \cite{smallbatch1}, and training with multiple dataset passes, also called epochs, is theoretically optimal relative to single-pass training \cite{smallbatch3}. We incorporate small-batch regularization (SBR) to train our model and achieve better generalization performance while preventing overfitting on our limited-sample dataset. 

\subsubsection{Model Overview}
Putting everything together, the bidirectionality and differential design ideas are integrated to form an updated ``Mamba++'' layer utilized in the NeuroMamba model. This layer is described by taking some input feature $F$ from the last layer and performing the following
\begin{tcolorbox}
\centering
\textbf{Mamba++ Layer}
\begin{align}
\tilde{F}_F &= \text{Mamba}(F), \quad \tilde{F}_B = \mathcal{T}(\text{Mamba}(\mathcal{T}(F))) \nonumber \\
F_O &= \lambda^2_{\theta} \odot \text{RMSNorm}(\lambda^1_{\theta} \odot \tilde{F}_F - \tilde{F}_B) + F
\end{align}
\end{tcolorbox}
\noindent where $\lambda^1_{\theta},\lambda^2_{\theta} \in \mathbb{R}^B$ are learnable scaling parameters, $\text{RMSNorm}(\cdot)$ is root mean square normalization \cite{rmsnorm}, and $\mathcal{T}(\cdot)$ is the linear operator for time flipping. Once the sequence has been processed by the Mamba++ layers, it is temporally averaged to summarize session statistics for every brain region. This forms a latent vector $h \in \mathbb{R}^B$ that is fed to an affine layer for score prediction. Let $f_{\theta}(\cdot)$ denote the NeuroMamba backbone that consists of a stack of Mamba++ layers combined with the temporal average pooling linear operator $\mathcal{P}(\cdot)$, $g_{\theta}(\cdot)$ denote the readout function that is simply an affine layer for linear regression, and $\theta$ correspond to the set of learnable parameters. Then, the NeuroMamba model is summarized by
\begin{tcolorbox}
\centering
\textbf{NeuroMamba}
\begin{align}
&f_{\theta}(\cdot) = \mathcal{P}([\text{Mamba++}_{\theta}^{(l)}(\cdot) \text{ for } l = 1 \dots L]) \\
&g_{\theta}(\cdot) = W_{\theta} f_{\theta}(\cdot) + b_{\theta} \\
&\min_{\theta} \sum_{i=1}^N \frac{1}{2} \underbrace{\| s - g_{\theta}(f_{\theta}(X^i))\|_2^2}_{\hidewidth \text{data fidelity} \hidewidth} + \lambda \underbrace{\| f_{\theta}(X^i) \|_1}_{\hidewidth \text{sparse brain regions} \hidewidth} \label{eq:neuromamba_cost}
\end{align}
\end{tcolorbox}
\noindent where \eref{eq:neuromamba_cost} is the cost function used to find optimal model parameters, $s$ contains the behavior score metrics, and $\| \cdot \|_1$ is an L1 penalty term to promote a sparse number of brain regions that are relevant for prediction. See \fref{fig:mambamodel} for a pictorial representation of the NeuroMamba model. The code associated with all of these methods and experiments in this paper is published at \href{https://github.com/javiersc1/NeuroMamba}{github.com/javiersc1/NeuroMamba}. In \sref{sec:results}, the experimental setup and results of these methods are discussed.

\section{Results \& Discussion}
\label{sec:results}

\subsection{Setup}

Due to the limited size of our dataset and the need to ensure generalization across the full distribution, we adopt a leave-one-out approach. Specifically, for each fold, the methods are trained on all data except one sample, which is held out for testing. This practice is commonly used in related fMRI prediction studies \cite{michelle_thesis, mocarsfmri, mocadepression}. The FCM, CPM, I-ICA, G-ICA, and ALFF algorithms perform feature extraction by optimizing a convex cost function, and subsequently employ kernel ridge regression (KRR), which is also a convex optimization method. Consequently, these methods are run until the optimization process converges. In contrast, the deep learning approaches like TCN, BiLSTM, PatchTST, and NeuroMamba involve non-convex optimization and are therefore trained for a fixed number of 50 epochs, as further epochs yield minimal improvements in the cost function at this stage for all methods. The Adam optimizer is used for training with momentum coefficients ($\beta_1=0, \beta_2=0.95$) which makes it behave similarly to RMSProp. Additionally,
gradient clipping is used for the recurrent networks to prevent exploding gradients. All relevant hyperparameters for the various methods were determined via cross-validation using a randomly selected training/validation split of $50\%/50\%$, ensuring that the chosen parameters are representative of the overall dataset. To promote fair assessment of model generalizability, these parameters remain fixed across all folds and are not re-optimized for each individual fold. Further, out of distribution testing is reported in \sref{sec:ood}.

\subsection{Experiments}

\begin{table}
  \centering
  \resizebox{0.98\columnwidth}{!}{%
  \begin{tabular}{llll}
    \hline
    Method & MoCA & Memory & Language \\
    \hline
    FC & $0.07$ & $0.08$ & $0.10$ \\
    $\text{CPM}_{\text{pos}}$ & $0.06$ & $0.01$ & $0.08$ \\
    $\text{CPM}_{\text{neg}}$ & $0.05$ & $0.03$ & $0.11$ \\
    \hline
    I-ICA & $0.14^{*}$ & $0.09$ & $0.10^{*}$ \\
    G-ICA & $0.18^{**}$ & $0.12^{*}$ & $0.16^{**}$ \\
    ALFF & $0.20^{***}$ & $0.13^{*}$ & $0.18^{**}$ \\
    \hline
    TCN & $0.26^{***}$ & $0.16^{**}$ & $0.17^{**}$ \\
    BiLSTM & $0.19^{***}$ & $0.19^{***}$ & $0.18^{**}$ \\
    PatchTST & $0.28^{***}$ & $0.17^{**}$ & $0.19^{***}$ \\
    NeuroMamba & $\textbf{0.36}^{***}$ & $\textbf{0.24}^{***}$ & $\textbf{0.25}^{***}$ \\
    \hline
  \end{tabular}
  }
  \caption{Pearson correlation coefficients ($R$) and corresponding p-values ($p$) by score category for multiple methods applied to the MADRC rs-fMRI data. Asterisks denote statistical significance as follows: $* = p < 0.1$, $** = p < 0.01$, $*** = p < 0.001$. }
  \label{tab:pcc}
\end{table}

\begin{figure*}
\centering
\begin{subfigure}{0.80\textwidth}
    \includegraphics[width=\textwidth]{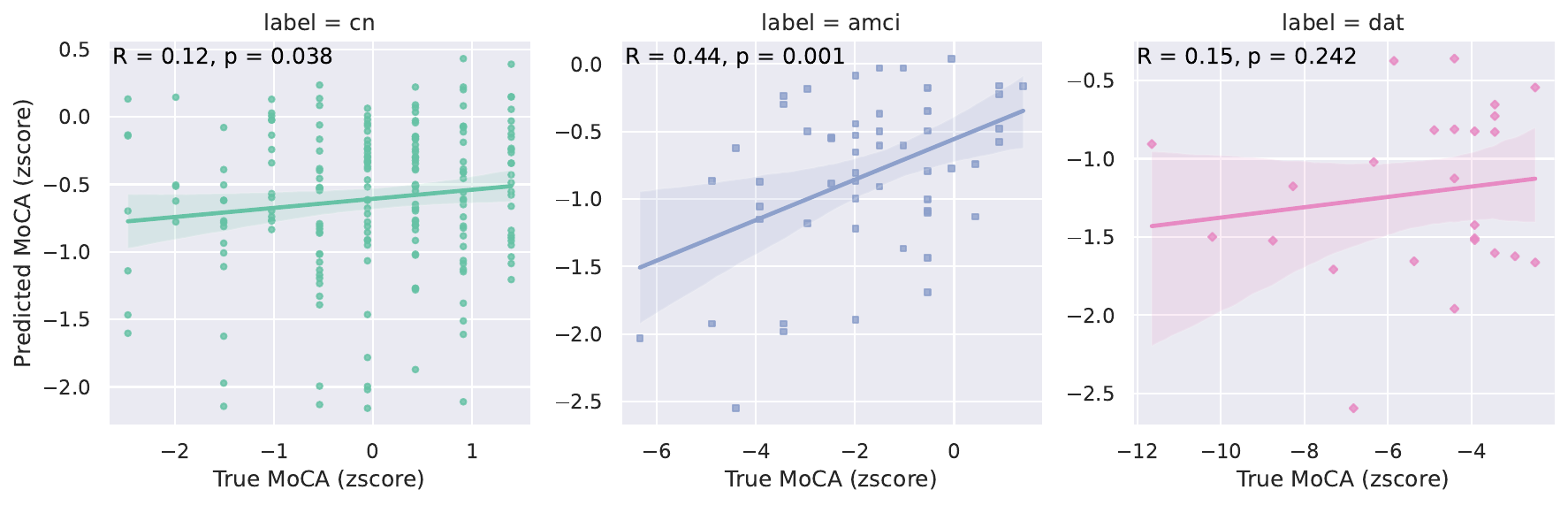}
\end{subfigure}
\hfill
\begin{subfigure}{0.80\textwidth}
    \includegraphics[width=\textwidth]{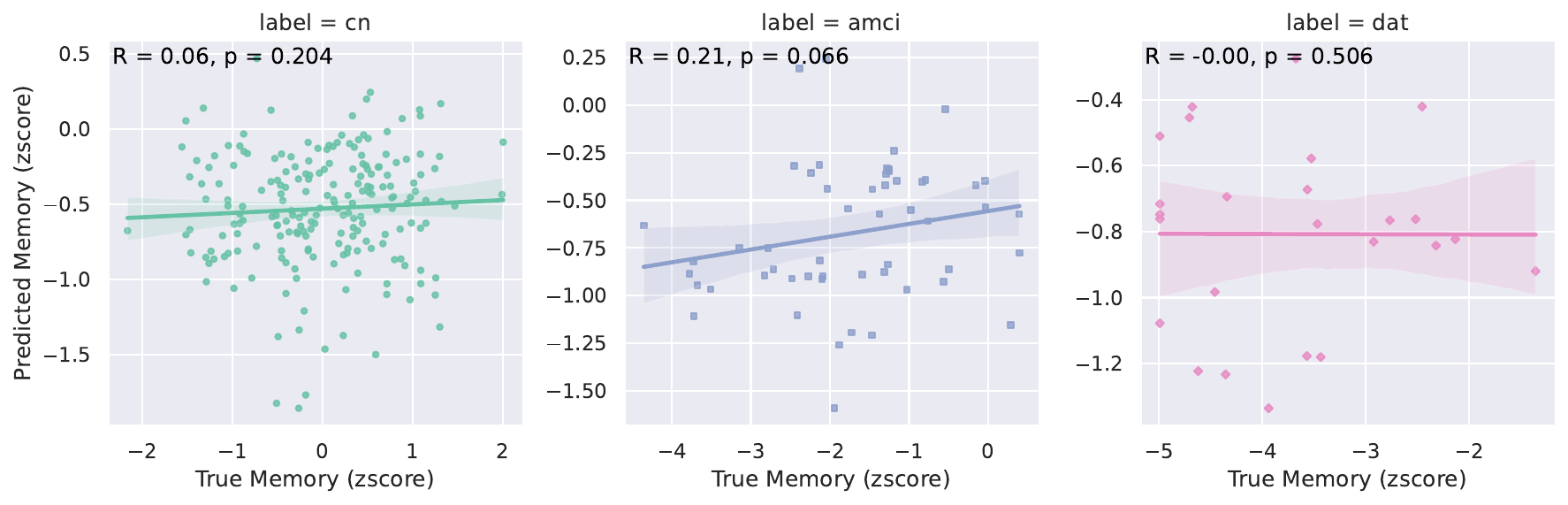}
\end{subfigure}
\hfill
\begin{subfigure}{0.80\textwidth}
    \includegraphics[width=\textwidth]{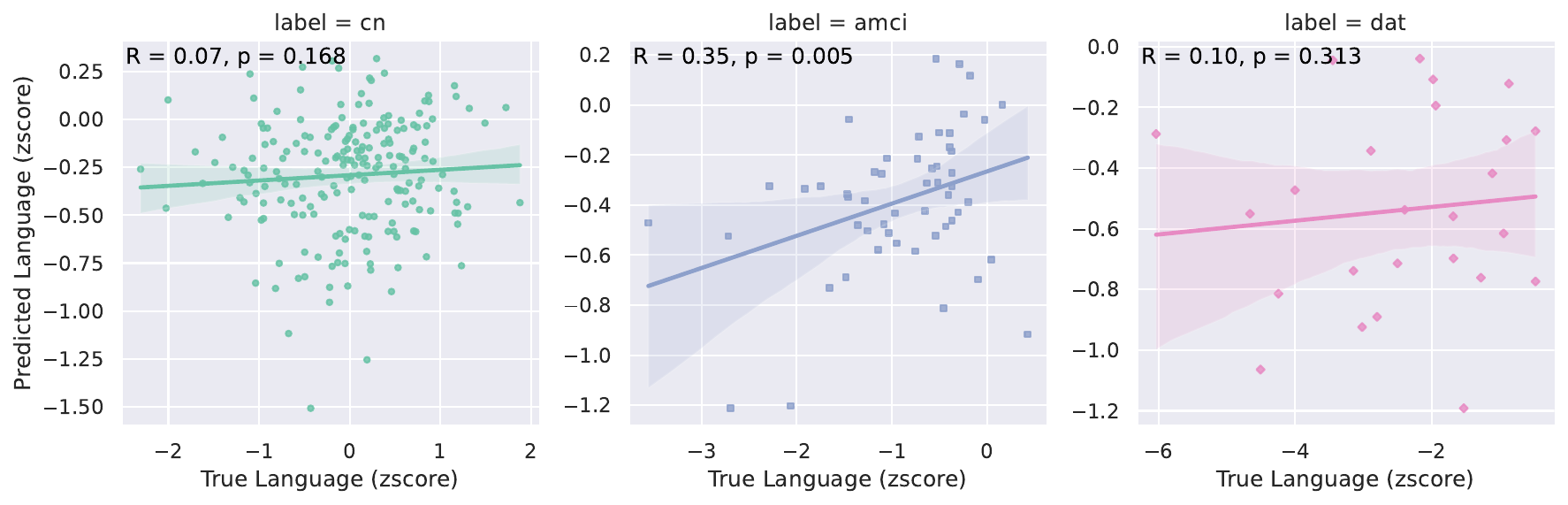}
\end{subfigure}
\caption{Correlation scatter plots displaying the relationship between predicted NeuroMamba scores and true behavioral metrics (rows), across the Alzheimer’s disease spectrum (columns), presented in z-score normalized space.}
\label{fig:scatter}
\end{figure*}

\subsubsection{Predictive Accuracy}
\label{sec:predictive_accuracy}
In this experiment, the accuracy of different approaches is compared by analyzing the Pearson correlation coefficient ($R$) and associated $p$-values between the predicted and true behavior scores for each cognitive subcategory. In \tabref{tab:pcc}, the $R$ values are reported for MoCA, average memory, and average language for each method; for brevity, only the MoCA $ R$ values are discussed below. As demonstrated, the functional connectivity method (FCM) achieves $R=0.07$, which closely aligns with published literature values of $R=0.07$ \cite{michelle_thesis} and $R=0.15$ \cite{mocadepression}. This is not a fair comparison given the different datasets and methods employed, yet our value appears in line with these works. Further, the CPM method achieves a similar range with the positive edge model achieving a value of $R=0.06$, which is not too different from FC. 

Regardless, this is where the literature ends, with connectivity-based methods. Compared to the timeseries approaches that perform better, this indicates that important temporal dynamics useful for behavior score prediction are lost. Shifting to model-based timeseries methods, I-ICA improves over the FCM baseline by learning the distinct sources that compose the BOLD data, leading to a MoCA value of $R=0.14$. However, since the sources are independent across subjects, this individual approach lacks group-level information that would enable shared sources among individuals. Thus, the G-ICA method with $R=0.18$ provides an advantage in finding shared sources at the population level. Yet, the ALFF approach turns out to be the best \emph{model-based} method, with $R=0.20$. 

However, these previous approaches use \emph{model-based} feature extraction that may not be optimally relevant for behavior score prediction. By comparison, the deep learning methods performed relatively well with values within $R=0.19-0.28$. Shifting to NeuroMamba, our \emph{data-driven} multivariate timeseries method, it achieved the highest value $R=0.36$, indicating superior capability in learning temporal dynamics in contrast to the reported CNN (TCN), Transformer (PatchTST), and RNN (BiLSTM) models.

\begin{table}
  \centering
  \resizebox{0.98\columnwidth}{!}{%
  \begin{tabular}{llll}
    \hline
    Method & MoCA & Memory & Language \\ \hline
    Mamba & $0.23^{***}$ & $0.12^{*}$ & $0.19^{***}$ \\
    \hspace{1mm} +bidirectionality & $0.28^{***}$ & $0.18^{**}$ & $0.19^{***}$ \\
    \hspace{1mm} +differential & $0.34^{***}$ & $0.24^{***}$ & $0.22^{***}$ \\
    \hspace{1mm} +SBR = NeuroMamba & $\textbf{0.36}^{***}$ & $\textbf{0.24}^{***}$ & $\textbf{0.25}^{***}$ \\
    \hline
  \end{tabular}
  }
  \caption{Comparative ablation analysis illustrating the performance differences between NeuroMamba and the standard Mamba architecture.}
  \label{tab:ablation}
\end{table}

\subsubsection{Ablation Study}
The ablation analysis shown in \tabref{tab:ablation} compares NeuroMamba with the standard Mamba architecture. These results reveal progressive improvements in predictive accuracy across cognitive categories. The bidirectional component with non-causal design leads to a $22\%$ improvement for MoCA relative to Mamba by better capturing long-range dependencies and patterns in the complex data. Further, the differential component leads to a $11\%$ improvement over the bidirectional Mamba variant by amplifying ``attention'' to related brain regions relevant to prediction. Finally, using small batch regularization (SBR) as a training technique leads to a $10\%$ improvement over the differential variant, demonstrating its use to prevent overfitting by introducing noisy gradient updates that improve generalization.

\subsubsection{Behavior Score Plots}
The predicted behavior scores are computed for the entire dataset and plotted against the true behavior scores in \fref{fig:scatter} for the NeuroMamba model with reported $(R,p)$-values for each behavior score and disease categories. Across the behavior scores, the highest correlation values are observed in the amnestic MCI group, suggesting its potential as a marker for at-risk populations that may develop into the DAT variety.

\begin{table*}
  \centering
  \resizebox{0.98\textwidth}{!}{%
\begin{tabular}{ccllll}
\toprule
PFI (RMSE) & Power ROI & MNI (x,y,z) & Nominal System & Lobe/Area & Talairach Daemon Label \\
\midrule
\midrule
\multicolumn{6}{c}{MoCA} \\
\midrule
$1.84 \pm 0.05$ & 77 & (-13, -40, 1) & Default Mode & Limbic Cortex & Parahippocampal Gyrus \\
$0.39 \pm 0.02$ & 170 & (6, -81, 6) & Visual & Occipital Lobe & Cuneus \\
$0.21 \pm 0.02$ & 177 & (-53, -49, 43) & Fronto-parietal Task Control & Parietal Lobe & Inferior Parietal Lobule \\
$0.07 \pm 0.01$ & 212 & (-11, 26, 25) & Salience & Limbic Cortex & Anterior Cingulate \\
$0.05 \pm 0.01$ & 93 & (15, -63, 26) & Default Mode & Occipital Lobe & Precuneus \\
\midrule
\multicolumn{6}{c}{Memory} \\
\midrule
$1.62 \pm 0.05$ & 77 & (-13, -40, 1) & Default Mode & Limbic Cortex & Parahippocampal Gyrus \\
$0.27 \pm 0.02$ & 177 & (-53, -49, 43) & Fronto-parietal Task Control & Parietal Lobe & Inferior Parietal Lobule \\
$0.26 \pm 0.01$ & 111 & (-11, 45, 8) & Default Mode & Limbic Cortex & Anterior Cingulate \\
$0.22 \pm 0.01$ & 170 & (6, -81, 6) & Visual & Occipital Lobe & Cuneus \\
$0.15 \pm 0.01$ & 93 & (15, -63, 26) & Default Mode & Occipital Lobe & Precuneus \\
\midrule
\multicolumn{6}{c}{Language} \\
\midrule
$0.79 \pm 0.02$ & 77 & (-13, -40, 1) & Default Mode & Limbic Cortex & Parahippocampal Gyrus \\
$0.60 \pm 0.01$ & 221 & (2, -24, 30) & Emotion/Behavior & Limbic Cortex & Cingulate Gyrus \\
$0.30 \pm 0.01$ & 170 & (6, -81, 6) & Visual & Occipital Lobe & Cuneus \\
$0.26 \pm 0.01$ & 177 & (-53, -49, 43) & Fronto-parietal Task Control & Parietal Lobe & Inferior Parietal Lobule \\
$0.26 \pm 0.01$ & 19 & (13, -33, 75) & Sensory/Motor & Frontal Lobe & Precentral Gyrus \\
\bottomrule
\end{tabular}
}
  \caption{Top five brain regions implicated in behavior score prediction, ranked by importance using permutation feature importance (PFI) for each score category. Additional columns provide MNI coordinates, nominal system category, lobe classification, and Talairach Daemon (TD) labels.}
  \label{tab:regions}
\end{table*}

\subsubsection{Impactful Brain Regions}
\label{sec:results_regions}
The NeuroMamba backbone $\mathcal{P}(f_{\theta}(X^i)) = h^i \in \mathbb{R}^{B}$ extracts $B$ elements, one element for each brain region in the list of Power atlas ROIs. This is fed to an affine layer $g_{\theta}(h^i) = W_{\theta} h^i + b_{\theta}$ to pick out some sparse combination of brain regions that are useful for prediction. In a standard linear regression problem where all features are z-score normalized, one can use the learned magnitude weights in $W$ to directly rank the most impactful features. However, in the deep learning context, this condition is not necessarily met, necessitating more advanced approaches to identify impactful features. 

In this work, permutation feature importance (PFI) \cite{fpi} is used to aid in this problem. PFI is a method for assessing the significance of features in a black-box model by measuring how much the prediction error increases when the values of a single feature are randomly shuffled across samples. If shuffling a feature significantly reduces the model's accuracy, that feature is considered important for the prediction task. This is done for each element $h \in \mathbb{R}^B$ independently across $100$ shuffling trials, while holding the other features fixed. The PFI method is applied for each behavior score category to identify relevant brain regions. The root mean square error (RMSE) metric is used to measure model performance degradation, and the top 5 brain regions for each behavior score are shown in \tabref{tab:regions}. Additionally, the Power atlas ROI index, MNI-space coordinates, nominal system, lobe/area, and Talairach Daemon labels \cite{td_labels} are included for each brain region to enhance discussion.

For brevity, only MoCA-related brain regions are discussed here. The top 5 brain regions selected are the parahippocampal gyrus, cuneus, inferior parietal lobule, anterior cingulate, and precuneus. The parahippocampal gyrus is important for memory encoding, retrieval, spatial navigation, and contextual processing \cite{hippo0}. For AD subjects, the parahippocampal gyrus shows atrophy with decline in episodic memory \cite{hippo1}, reduced activation during memory tasks in fMRI \cite{hippo2}, early deposition of amyloid-beta plaques and tau proteins \cite{hippo3}, and is considered a key node in the DMN with disrupted connectivity \cite{hippo4}. Because of this, it is not surprising that this region had a significant impact. The cuneus, a region of the occipital lobe responsible for higher-order visual processing, has been implicated in cognitive impairment through atrophy and hypometabolism associated with deficits in visual processing, spatial navigation, and attentional functions in AD subjects \cite{cuneus1, cuneus2, cuneus3}. The precuneus is a region with many daydreaming-like functions, such as mental imagery, episodic memory, and self-reflection, and it is an active part of the DMN in rs-fMRI \cite{precuneus0}. Similarly, the precuneus shows reduced metabolism and atrophy in AD subjects \cite{precuneus1, precuneus2}. The inferior parietal lobule (IPL) is involved in various cognitive processes, including speech, language, spatial reasoning, working memory, and number processing \cite{ipl1}. The IPL undergoes changes in thickness of its banks during the transition from healthy to mild impairment \cite{ipl2}, and abnormal functional connectivity changes occur relative to other networks, such as salience, sensorimotor, and executive \cite{ipl3}. Lastly, the anterior cingulate (ACC) plays a major role in emotion regulation and processing, like impulse control, motivation, goal-directed behavior, and emotional pain perception \cite{acc1}. For AD subjects, the ACC exhibits reduced thickness \cite{acc2} and decreased functional connectivity \cite{acc3}, which has connections to high agitation, irritability, and anxiety in people with AD \cite{acc4}. To summarize, our findings of relevant brain regions for behavior score prediction in rs-fMRI closely align with established medical research on impaired regions in AD, which may be useful for intervention strategies such as multi-region HD-tDCS applications.

\begin{figure}
    \centering
    \includegraphics[width=0.9\columnwidth]{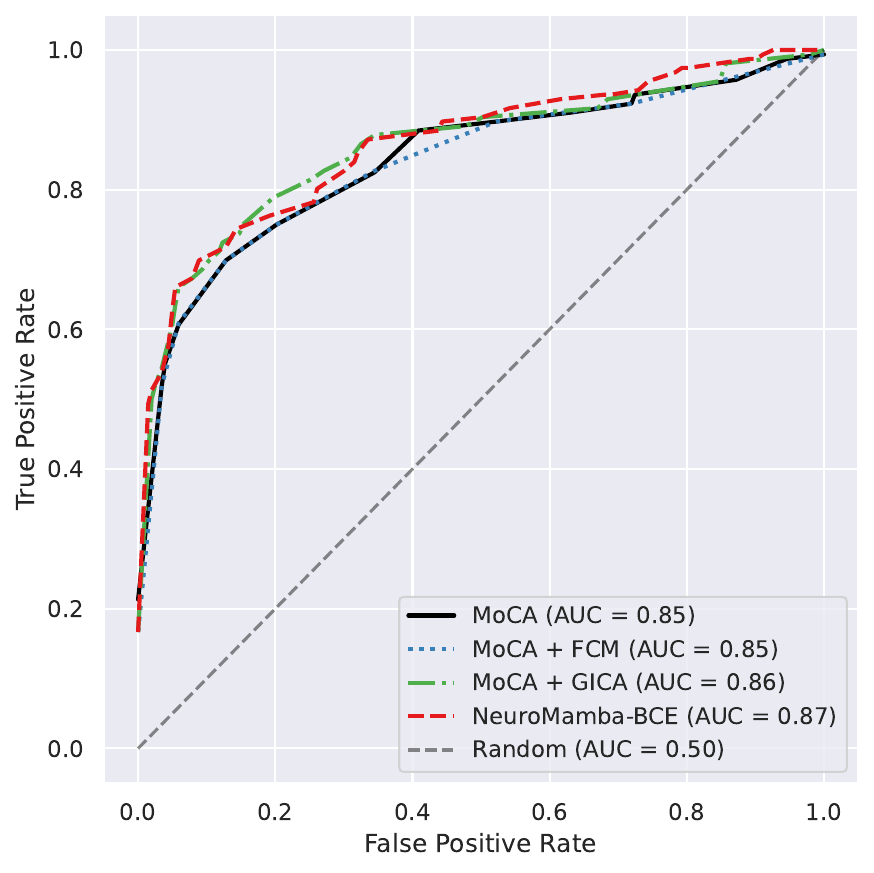}
    \caption{Receiver operating characteristic (ROC) curve with area under curve (AUC) values for diagnosis between cognitively normal and non-normal subjects.  }
    \label{fig:roc}
\end{figure}

\subsubsection{Early Biomarker Feasibility}
There have been many works that explore classification/diagnosis of subjects in the AD spectrum using rs-fMRI data \cite{classification1, classification2, classification3, classification4, classification5, classification6}. They all follow the same general formula: 1) introduce the idea that functional changes occur in the brain earlier than structural ones meaning there is promise for early diagnosis, 2) develop some novelty in the methodology, e.g., going from a convolutional to a transformer-based network, that performs relatively better for diagnosis, and 3) reiterate that these results indicate the potential for rs-fMRI data to be used as an early biomarker for diagnosis. 
However, these works and many others miss the mark. The more medically relevant question is whether rs-fMRI data adds something useful to diagnosis that is not already provided by easily acquired metrics such as MoCA. In this experiment, we aim to analyze the classification performance of only using MoCA, MoCA with FCM features, MoCA with G-ICA features, and MoCA with the proposed NeuroMamba model. For the model-based approaches, logistic regression is used to predict the probability of belonging in the positive class (aMCI and DAT) from the negative class (CN). For NeuroMamba, a variant named NeuroMamba-BCE is created that combines the backbone model $f_{\theta}$ with MoCA values $s$ by
\begin{tcolorbox}
\centering
\textbf{NeuroMamba-BCE}
\begin{align}
&f_{\theta}(\cdot) = \mathcal{P}([\text{Mamba++}_{\theta}^{(l)}(\cdot) \text{ for } l = 1 \dots L]) \\
&g_{\theta}(\cdot, s^i) = W_{\theta} \hspace{1mm} \text{Concat}[\underbrace{f_{\theta}(\cdot)}_{\hidewidth \text{rs-fMRI features} \hidewidth}; \hspace{1mm} \overbrace{s^i}^{\hidewidth \text{MoCA} \hidewidth}] + b_{\theta} \\
&\min_{\theta} \sum_{i=1}^N \underbrace{ \mathcal{L}_{\text{BCE}}( g_{\theta}(X^i , \hspace{1mm} s^i); \hspace{1mm} c^i ) }_{\hidewidth \text{binary cross entropy} \hidewidth} + \lambda \underbrace{\| f_{\theta}(X^i) \|_1}_{\hidewidth \text{sparsity reg.} \hidewidth} 
\end{align}
\end{tcolorbox}
\noindent and train the model using the binary cross entropy loss function to learn the mapping between features $\{X^i, s^i\}$ and subject label $c^i$. Since it directly predicts probabilities, logistic regression is not needed here in this deep approach. 

The receiver operating characteristic (ROC) curves for these approaches are plotted in \fref{fig:roc}, along with the corresponding area under the curve (AUC) values. From these results, there was no improvement in diagnostic ability across connectivity and time series based methods, including our proposed model. This negative result indicates that \emph{resting-state} fMRI is unlikely to provide additional diagnostic power beyond that already provided by MoCA and other readily available metrics. However, it is possible that \emph{task-based} fMRI could provide more complementary information, given it is more likely to better ``stress'' the brain networks. Further research is necessary in this underexplored area to conclusively determine whether fMRI as a whole is beneficial for early diagnosis.

\subsubsection{Out of Distribution (OOD) Generalization}
\label{sec:ood}
The Alzheimer's Disease Neuroimaging Initiative (ADNI) \cite{adni} began in 2004, with the main goals of providing data to researchers and improving doctors' diagnoses of subjects with AD. Part of this data collection includes the MoCA exam and 3T rs-fMRI EPI data (Flip Angle=$90^{\circ}$, TE=30ms, TR=3.0s, 3.4mm spatial resolution, 192 total time samples), which gives us an opportunity to explore how these methods generalize to a different dataset. The ADNI dataset is prepared using a methodology similar to that in \sref{sec:preprocess}, with a few key differences. Namely, fieldmap data is inconsistent and missing for some subjects, so fieldmap-less susceptibility distortion correction is done instead. Additionally, since the TR is quite high, slice timing correction is needed to ensure that all slices are treated as if they were acquired simultaneously. The MoCA scores are z-score normalized per the MADRC normal population. The ADNI data consists of 471 subjects (324 CN, 110 aMCI, 37 DAT) over 860 sessions (579 CN, 220 aMCI, 61 DAT). Care is taken to ensure no leakage during testing, i.e., a subject's scans are held out entirely rather than only a single session.  

In this section, we test for two things: zero-shot transfer and all-shot training. Zero-shot means a method is trained on MADRC and tested on ADNI using the same hyperparameters from MADRC. All-shot means trained on ADNI directly using the same hyperparameters from MADRC. The ADNI data contains 192 time samples with a $\text{TR}=3.0s$, whereas MADRC data contains 570 time samples with a $\text{TR}=0.8s$. Therefore, this section explores the effect of temporal resolution on the predictive accuracy of MoCA and other key differences between the two data distributions. Similar to \sref{sec:predictive_accuracy}, \tabref{tab:adni} shows MoCA R values for the ADNI data under the zero-shot and all-shot settings. 

Functional connectivity appears unaffected in both scenarios, likely because it is independent of temporal dynamics. Notably, the CPM positive edge method showed a higher correlation value $R=0.23$ on all-shot in contrast to zero-shot $R=0.07$, indicating some distribution shift between MADRC and ADNI CPM-derived brain scores. The model-based timeseries methods did not exhibit strong zero-shot transfer or all-shot performance, with G-ICA being the notable exception, indicating that the ideal hyperparameters are a function of temporal resolution. Switching to data-driven timeseries methods, PatchTST and TCN showed weak zero-shot and all-shot performance, likely because the ideal patch size depends on the temporal resolution and kernel sizes. The NeuroMamba model exhibited the strongest zero-shot transfer among all methods, with $R=0.17$, and showed the most robustness to temporal sampling effects, matching literature observations comparing SSMs to other models \cite{s4}. When trained directly on ADNI, NeuroMamba achieved the highest all-shot performance ($R=0.36$), indicating strong generalizability to other datasets. 

\begin{table}
  \centering
  \resizebox{0.75\columnwidth}{!}{%
  \begin{tabular}{l|l|l}
    Method & Zero-shot & All-shot  \\ \hline
    FC & $0.12^{***}$ & $0.12^{***}$ \\
    $\text{CPM}_{\text{pos}}$ & $0.07^{*}$ & $0.23^{***}$ \\
    $\text{CPM}_{\text{neg}}$ & $0.05^{*}$ & $0.18^{***}$ \\
    \hline
    I-ICA & $0.04$ & $0.05^{*}$ \\
    G-ICA & $0.03$ & $0.29^{***}$ \\
    ALFF & $0.05^{*}$ & $0.01$ \\
    \hline
    TCN & $0.09^{**}$ & $0.16^{***}$ \\
    BiLSTM & $0.12^{***}$ & $0.29^{***}$ \\
    PatchTST & $0.08^{**}$ & $0.08^{**}$ \\
    NeuroMamba & $0.17^{***}$ & $0.36^{***}$ \\
  \end{tabular}
  }
  \caption{Out of domain (OOD) generalization on Alzheimer's Disease Neuroimaging Initiative (ADNI) dataset where values indicate MoCA Pearson correlation.  Asterisks denote statistical significance as follows: $* = p < 0.1$, $** = p < 0.01$, $*** = p < 0.001$. }
  \label{tab:adni}
\end{table}

\subsubsection{Domain Adaptation}
For NeuroMamba, per the results in \sref{sec:ood}, there is a gap between zero-shot $R=0.17$ and all-shot $R=0.36$ values. This provides an opportunity to study whether it is possible to close the gap by learning on a tiny subset of ADNI data, using the pretrained MADRC model to adjust for any distribution mismatch. In the literature, there are many ways to do this, such as finetuning, freezing the backbone and updating only the last few layers, or using deep learning approaches such as correlation alignment (CORAL) \cite{coral}, designed to minimize covariance differences between domains. In this setting, we found great success in simply finetuning the pretrained model. As observed in \tabref{tab:fewshot}, with only 5 subjects per class for training data, NeuroMamba achieves an $R=0.35$ value, nearly matching all-shot performance. This finding demonstrates the model's ability to adapt to a different domain with very limited training data, which is highly practical for many fMRI-related applications.

\begin{table}
  \centering
  \resizebox{0.60\columnwidth}{!}{%
  \begin{tabular}{l|l}
    NeuroMamba & MoCA \\ \hline
    Zero-shot & $0.17^{***}$ \\
    Single-shot & $0.22^{***}$ \\
    Three-shot & $0.30^{***}$ \\
    Five-shot & $0.35^{***}$ \\
    All-shot & $0.36^{***}$
  \end{tabular}
  }
  \caption{Domain adaptation on Alzheimer's Disease Neuroimaging Initiative (ADNI) dataset where values indicate MoCA Pearson correlation.  Asterisks denote statistical significance as follows: $* = p < 0.1$, $** = p < 0.01$, $*** = p < 0.001$. }
  \label{tab:fewshot}
\end{table}
\section{Conclusion}
\label{sec:conclusion}
In this work, we sought to advance understanding and prediction of cognitive performance in AD by leveraging rs-fMRI data alongside behavioral scores, including MoCA, average memory, and average language metrics. By systematically evaluating the predictive power of both functional connectivity and multivariate timeseries data, we addressed limitations in prior studies that focused exclusively on functional connectivity. Our deep learning approach, based on state space modeling, demonstrated superior performance in predicting behavioral metrics, underscoring the value of temporal dynamics in rs-fMRI data. Furthermore, NeuroMamba achieved a remarkable few-shot transfer performance on ADNI data, indicating that only a few subjects are needed when finetuning the MADRC-pretrained model on out-of-domain datasets such as ADNI, which is important for real-world usability. These results highlight the potential to integrate machine learning with neuroimaging to support early intervention for cognitive decline, using techniques such as HD-tDCS, while simultaneously offering deeper biological insights into AD progression. 

However, this proposed approach has limitations. The R values across the behavior scores are modest, indicating possible limitations with using \emph{resting-state} fMRI to analyze cognition scores. It is likely that \emph{task-based} fMRI, in which a patient performs a task under the scanner, could better ``stress'' brain networks, revealing a stronger relationship between key regions and behavioral metrics. For future work, one can explore face-name association \cite{facename} and object-location association \cite{objectlocation} tasks within a \emph{task-based} functional MRI framework to compare with the findings in this work and gain further insights by systematically contrasting resting-state and task-based functional activity.

\bibliographystyle{IEEEtran}
\bibliography{refs.bib}

\end{document}